\documentclass[aps,preprint]{revtex4}
\usepackage{epsfig}
\newcommand{\Deg}{{}^{\,\mbox{\scriptsize o}}}          % -  degree

\begin{document}
%% for editor

\title{\large
Ultra High Energy $\nu_\tau$ Detection Using Cosmic Ray Tau Neutrino Telescope\\ 
Used in Fluorescence/Cerenkov Light Detection}

\author{\bf Z.\,Cao$^{1,2}$, M.\,A.\,Huang$^3$,  P.\,Sokolsky$^2$, Y. Hu$^4$} 
\affiliation{\it $^1$ Institute of High Energy Physics, Beijing, 100039, China\\ 
$^2$ High Energy Astrophys. Inst., Univ. of Utah, Salt Lake City UT 84112 USA\\
$^3$ General Education Center, National United University, Miaoli, 360, Taiwan 
R.O.C.\\$^4$ University of Science \& Technology of China, Hefei, 230026, China} 

\begin{abstract}
We have investigated the possibility of $\nu_\tau$ detection using Cosmic
Ray Tau Neutrino Telescope (CRTNT) based on air shower 
fluorescence/Cerenkov light detector techniques. This approach requires an 
interaction of a $\nu_\tau$ with material such as a mountain or the earth's crust. $\tau$ 
lepton produced in the charged current interaction must escape from the earth and 
then decay and initiate a shower in the air. The 
probability for the conversion from $\nu_\tau$ to air shower has been calculated for 
an energy range from 1 PeV to 10 EeV.  
An air shower simulation program has been developed using the simulation package Corsika.  
The trigger efficiency has been estimated for a CRTNT detector similar 
to the HiRes/Dice detector in the shadow of Mt. Wheeler in 
Nevada, USA. A rate of about 8 triggered events per year is expected for the AGN 
neutrino source model with an optimized configuration and duty cycle of 
the detector. 

\end{abstract}

\pacs{96.40.Tv, 96.40.Pq, 98.70.Sa, 95.55.Vj}
\maketitle

\section{Introduction}
The source of cosmic rays with particle energies above $10^{15}$ eV (1 PeV) 
remains unknown. A point source search could be an encouraging approach for 
solving this puzzle. Searches for point sources are best performed using 
observations of neutral particles because they can be directly traced back to 
the source. The universe is opaque to photons between $10^{14}$ eV and at 
least $10^{18}$ eV (1 EeV) due to interaction with the 2.7$\Deg$ K 
cosmological microwave background. The neutrino is another type of neutral 
particle that can be used to explore cosmic ray sources in this energy region. 
Newly discovered evidence of neutrino oscillation~\cite{nuoscill} leads to a 
plausible argument that the astrophysics neutrino flux will have an even 
flavor ratio of $\nu_e:\nu_\mu:\nu_\tau=1:1:1$~\cite{NU-MIX}. 
%%% Added by Alfred
One obvious effect of neutrino mixing is the appearance of tau 
leptons. This opens a window for the cosmic ray source 
searches using air shower techniques since the decay products of $\tau$ lepton, mainly 
hadrons and electrons, will induce a detectable shower in the air. 

 Neutrinos convert to electrons, muons 
and taus through the charged current interaction. 
The interaction probability is much higher in the Earth than in the atmosphere, 
due to the higher density of rock. However, electrons will shower quickly inside the target 
material. Muons travel very long distances before they decay but can only be 
detected by the small energy loss along the trajectory. A large detector, such as 
IceCube, buried inside the target material, is designed to detect those muons 
(track-like signals) and electrons (shower-like signals), but is inefficient 
for tau detection. At energies below 1 PeV, the tau decays and initiates 
a shower that can not be distinguished from an electron shower. 
At energies above 20 PeV, the tau decay distance is longer than the width of 1 km$^3$ 
size detector and only one shower can be seen. At higher energies, the neutrino flux 
is much lower and a 1 km$^3$ size detector is not large enough to detect such a flux. 

One way to avoid constraints from the target volume is to separate of detection 
volume from the target volume. A fluorescence light detector such as HiRes or a 
Cerenkov light detector such as Dice has proven 
to be a successful detector with a small physical size but a huge detection volume. 
Leptons produced via charged current interaction
 must be able to escape from target volume, decay in the atmosphere 
and develop a shower before they reach  
the detector for this scheme to work. 
Since the Earth becomes essentially opaque to neutrinos at energy higher than 1 PeV, 
neutrinos come out in almost horizontal direction. These are
known as {\em earth-skimming neutrinos}~\cite{Feng, fargion}. The conversion 
efficiency from a tau neutrino to a tau is approximately $R_d \rho / \Lambda_{I}$, where 
$\rho$ is target density. $R_d$ is range of the tau in the target and 
$\Lambda_{I}$ is interaction length in the target~\cite{Alfred-1}. 
At energies above 1 PeV, tau leptons have 
a range such that the conversion probability in the Earth is higher than that 
in the atmosphere~\cite{Alfred-2}. The conversion 
efficiency becomes much higher at higher energy because $R_d$ 
increases less than linearly with energy but $\Lambda_{I}$ grows proportional to 
$E^{-2.36}$~\cite{Gandhi}. 
This unique mechanism of detecting $\tau$-neutrino provides an acceptance that is much 
larger than 1 km$^3$ underwater/under-ice detector arrays, especially at energy 
higher than 10 PeV. 
Detecting the tau lepton from earth-skimming neutrinos becomes an excellent way to 
study ultra high energy neutrino astronomy and to provide a possible proof of 
neutrino flavor oscillation. 

The rest of paper is organized as following. In 
Section II, the conversion from tau neutrino to air shower 
and possible shower detection are discussed. The CRTNT 
detector is proposed at a potential site. 
A simulation of  $\nu_\tau$ to $\tau$ conversion, 
shower development and shower detection 
is described in Section III in detail. The 
event rate and its optimization are presented in Section IV. 
Final conclusions are summarized in the last section.

\section{Feasibility study} 
We first try to evaluate the detection of tau neutrinos skimming 
through mountains, then 
discuss the detection techniques and a potential site for the proposed CRTNT project.

\subsection{ $\nu_\tau$ to air shower conversion} 
\begin{figure}
  \epsfig{file=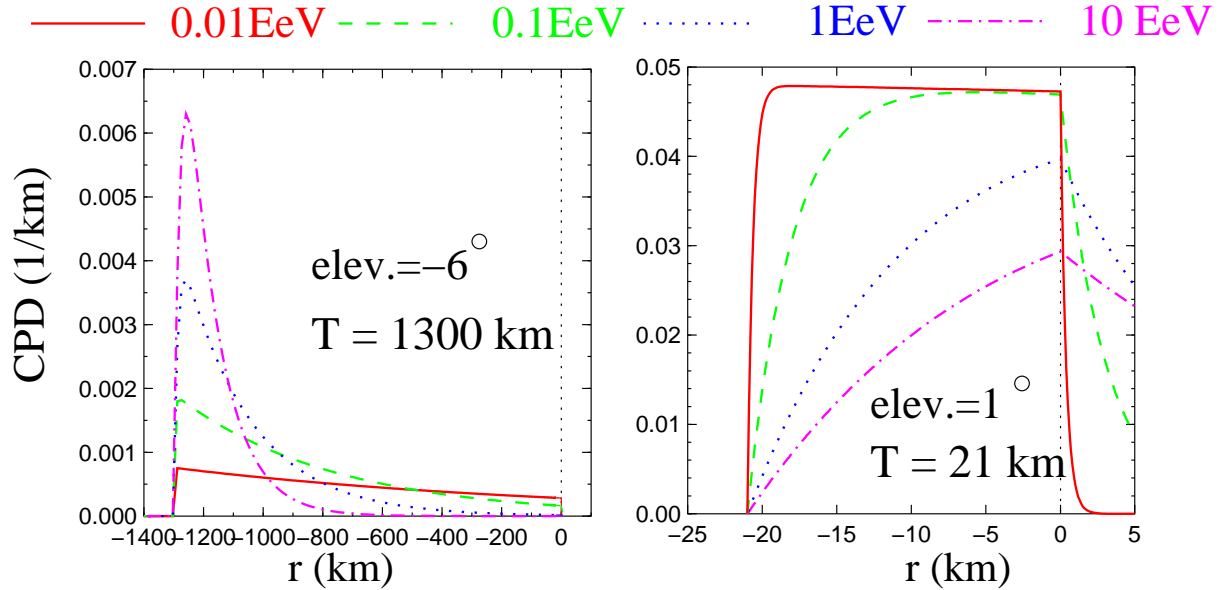,width = 16cm}
  \caption{$\nu_\tau$ to shower conversion probability density (CPD) as a 
function of distance along trajectory.}
  \label{cpd_x}
\end{figure}

Two coordinate systems are used in this study. A three dimensional local 
coordinate system describes the detector position and local 
topological information. 
The other is an one-dimensional coordinate system along the neutrino trajectory 
used for describing the mountain-passing/Earth-skimming process. 
The trajectory is described by the elevation angle 
$\epsilon$, the azimuth angle $\phi$, and the location of the escape point from the 
mountain/earth surface ($x_e,y_e,z_e$) in the 3-D local coordinate system. 
In the 1-D coordinate system, $\nu_\tau$'s travel from negative infinity and enter
 the mountain at $r=-T$, where T is the thickness of the mountain or the earth's crust. 
The density 
distribution $\rho(r)$, and the total path length $T$ are different depending on
the geometry of the trajectory. The incident $\nu_\tau$ 
interacts between $-T$ and 0, the $\tau$ then decays and initiates a shower at $r$. 
The probability of such a conversion from $\nu_\tau$ to a shower is a 
convolution of interaction probability of $\nu_\tau$ before distance $r$ and the 
probability of decay to a shower at $r$. It simplified to 
\begin{eqnarray}
 p_c(r; T)dr = \left\{ \begin{array}{ll}
 \frac{1}{1-e^{-T\rho/\Lambda}}\frac{dr}{R_d-\Lambda_I/\rho}\left[ \mbox{e}^{-
\frac{r+T}{R_d}} 
- \mbox{e}^{-\frac{r+T}{\Lambda_I/\rho}}\right] &\mbox{(-T$<$r$<$0)}\\
 \frac{1}{1-e^{-T\rho/\Lambda}}\frac{dr}{R_d-\Lambda_I/\rho}\left[ \mbox{e}^{-
\frac{T}{R_d}} 
- \mbox{e}^{-\frac{T}{\Lambda_I/\rho}}\right]\mbox{e}^{-\frac{r}{R_d}} 
&\mbox{(r$\geq$0)}
                  \end{array} \right.
\label{cpd}
\end{eqnarray}
where $p_c(r; T)$ is the conversion probability density (CPD) for a given 
trajectory $T=T(\epsilon,\phi,x_e,y_e)$, $\lambda_I$ is the $\nu_\tau$ interaction 
length in $g/cm^2$ and $R_d$ is the average range of $\tau$-lepton in km, i.e.
 $R_d=f \gamma c\tau$ where $\gamma$ is the Lorentz boost factor, 
 $\tau$ is the lifetime of the $\tau$-lepton and $f$ is a fraction accounting 
energy loss~\cite{duttaina}. $f$ decreases from 0.8 at $E_{\tau}=100$ PeV to 0.2 
$E_{\tau}=1$ EeV~\cite{Alfred-1}. FIG.~\ref{cpd_x} shows two cases of the 
CPD's as functions of $r$ corresponding to down-going ($\epsilon>0$, through a 
mountain about 21 km thick) and up-going earth-skimming ($\epsilon<0$, 
through the earth's crust about 1300 km thick) trajectories.

As shown in FIG. \ref{cpd_x}, for energies lower than 10 PeV, the CPD is flat 
because of the long interaction length of the $\nu_{\tau}$ and the short decay length 
of the $\tau$ lepton. Only primary $\nu_{\tau}$'s that interact close to the 
surface can produce an observable $\tau$. For $E_\nu$ above 10 PeV, the CPD peaks around 10-20 km 
in thickness of rock. The energy loss in the target material reduces both the 
survival rate and the energy of $\tau$'s above 10$^{17}$ eV. These results are 
consistent with other recent calculations~\cite{fargion, Alfred-1}.

\subsection{Detection of earth skimming $\nu_\tau$} 
The $\tau$ lepton flux induced by an earth-skimming neutrino flux will be orders 
of magnitudes lower than the cosmic ray flux in the observational window, a
range from 1 PeV to 10 EeV~\cite{Tseng}. A successful detection would require a 
detector having a large acceptance, on the order of several km$^2$ sr 
to over 100 km$^2$ sr. Because most of earth-skimming neutrinos are 
concentrated near the horizon, the detector should have a great sensitivity to 
horizontal showers. Ground detector arrays have small trigger efficiency for both 
upward and downward going showers in the vicinity of horizontal direction and 
hence can not be very efficient for detecting $\nu_\tau$. 
Detection through optical signals would seem to be more efficient. A combination
of fluorescence light and Cerenkov light detector in a shadow of a steep 
cliff could achieve this goal. 

% because the optical
%detector can have a wide field of view (FOV), especially a large azimuth angle 
%coverage and a large aperture.
A fluorescence light detector, such as the HiRes prototype 
experiment~\cite{HiRes_MIA}, has a large FOV and acceptance. The difficulty is 
that it has to be operated at a high threshold corresponding to above 
0.3 EeV, because of noise from sky background light. 
Cerenkov radiation provides many more photons along the shower axis. 
It can be useful for lowering the detection threshold. The Dice 
Experiment~\cite{Dice} successfully 
measured the energy spectrum and composition of cosmic rays between 0.1 PeV 
and 10 PeV using the same detector but 
triggered by Cerenkov light. To maintain a good energy resolution, a typical Cerenkov 
telescope has to be operated within a small FOV of about a few degrees with respect 
to the air shower since the Cerenkov light intensity is a rapidly decreasing 
function of the viewing angle. On the other hand, a fluorescence light 
detector has to be operated with large viewing angle to the shower axis to 
avoid significant Cerenkov light contamination that can cause poor shower energy 
resolution. 
In order to achieve satisfactory statistics for the neutrino flux measurement, 
we have to lower the threshold while maintain a large aperture. 

Several points need to be stressed.
First of all, the shower energy resolution is not crucial for neutrino 
searches, hence the Cerenkov light can be tolerated and can be useful for  
triggering the detector. Secondly, sky noise can be 
suppressed by placing the detector in the shadow of a mountain.
A steep mountain in front of the detector is shown to be a good target for neutrino 
conversion in the above discussion. It also serves as a screen to block sky noise. 
Because of the low light background, the detector can be operated 
with a lower threshold as a fluorescence light detector. More importantly, 
the background from cosmic rays at the same or lower energies is much reduced. 
 This makes the $\nu_\tau$ events distinguishable from cosmic ray events.
Finally, to enlarge the FOV of the detector, the telescopes can be configured 
to surround the mountain and minimize
overlaps between telescopes.  

In order to obtain sufficient fluorescence and Cerenkov light, we have to have 
a sufficient space for air shower development. A steep
mountain side provides a large elevation coverage for the detector and a large 
open space between the mountain surface and the detector.
An ideal site for observations would be by a steep mountain and in a
 dry and clear atmospheric environment essential
for fluorescence/Cerenkov light techniques. 

In this paper, we consider placing the CRTNT detector 
in proximity to Mt. Wheeler Peak(3984 m a.s.l.) 
near the Nevada-Utah border, USA. The mountain lies in a north-south 
direction and is about 40 km long. The west side is very steep. Further 
west, there is a flat valley about 30 km wide at about 1500 m a.s.l. 
For a detector located about 12 
km away from the peak horizontally, the shadow of the mountain is about 
11.7$^\circ$ in elevation. It almost completely blocks the FOV of the telescopes 
of the CRTNT detectors proposed below. Within the shadow, the sky noise 
background should be less 
than that from the open sky. At Dugway, Utah, about 120 km north 
of the Wheeler Peak, the light background was measured to be 40 
photon/$\mu$sec/m$^2$. In this paper, the light background for all the telescopes 
in the shadow is assumed to be the same as at Dugway. This is the upper limit 
to the sky noise. An on-site 
measurement for the actual 
background light will be done using a prototype detector. 

\subsection{The CRTNT fluorescence/Cerenkov light detector}
\label{detector}
The proposed CRTNT project uses fluorescence/Cerenkov light telescopes analogous to 
the detectors of the HiRes and the Dice experiments\cite{HRprototype}. 
The telescopes are distributed 
in three groups located at three sites separated by eight km in the valley 
facing the west side of Mt. Wheeler Peak. At each site, four telescopes observe an area of 
the mountain with 64$^\circ$ in azimuth and 14$^\circ$ in elevation. 
This area is about 14 km long in the north-south direction. The total field of 
view covers about $60^\circ\times 14^\circ$ and 70 km$^2$ with small overlaps
between the telescopes on the three sites.  

A 5.0 m$^2$ light collecting mirror with an reflectivity of 82\% is used for 
each telescope. The focal plane camera is made of 16$\times$16 pixels. 
Each pixel is a 44 mm hexagonal photomultiplier tube that has about a
$1^\circ\times 1^\circ$ field of view. 
Each tube is read out by 50 MHz flash ADC electronics system to measure the waveform 
of the shower signals
 
A pulse area finding algorithm is developed for providing an
individual channel trigger using field programmable gate array (FPGA). 
The first level trigger is set by requiring the signal-noise ratio to be greater than
4$\sigma$, where the $\sigma$ is the standard deviation of the total
photo-electron noise during the signal pulse. The noise level is different
for each pulse because the pulse duration varies depending on the distance
to the light source.
 The second level trigger requires at
lease five channels triggered within a 5$\times$5 running box over a single
telescope camera of 16$\times$16 pixels. The trigger condition for an event is that at least
one telescope is triggered. All triggers are formed by FPGA chips. Event 
data from all channels are scanned with a threshold lower than the trigger threshold
from the FPGA buffers into a local Linux box.  
 
\begin{figure}
  \epsfig{file=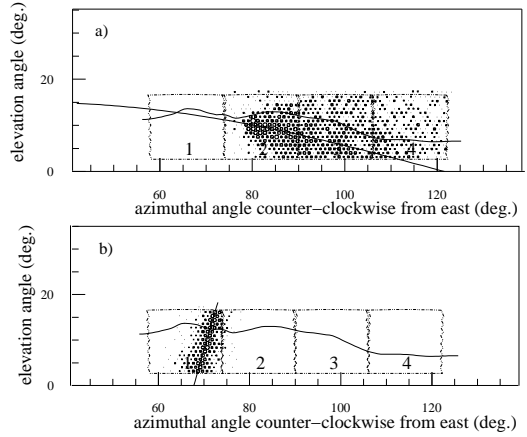,width = 7cm}
  \caption{Simulated air showers seen by the CRTNT detector. a) A $\tau$ neutrino
induced air shower starting and developing completely in the shadow of the mountain. 
The line along the shower direction represents a plane which contains the detector 
and the shower axis. The dashed lines show the boundary of the field of view of 
each telescope. The thick solid curve represents the profile of the Mt. Wheeler Peak. 
Circles represent triggered tubes and the size of each circle is proportional 
to the logarithm of the number of photons seen by the tube. b) A normal cosmic ray 
air shower event coming down and hitting on the slope of the mountain. }
  \label{events}
\end{figure}

A Monte Carlo simulation program for the detector 
is developed as described in the next section. Thousands of showers 
initiated by the products of $\tau$-decays and normal cosmic rays are generated.
 Two simulated event examples are shown in 
FIG.\ref{events}. In FIG.\ref{events} a), a $\tau$
neutrino induced air
shower starts in the shadow of the mountain where a normal cosmic ray event is
not expected. By contrast, a normal cosmic ray event is shown in FIG.\ref{events} b).
FIG.\ref{crtnt1} shows the detector configuration and shower initial point locations.

\section{Monte Carlo simulation}
In the simulation, the incident $\nu_\tau$ is coming from an interval of
elevation angles between $-11^\circ$ and 17$^\circ$, where the negative 
direction means a upward-going neutrino. The azimuth range is from
90$^\circ$ to 270$^\circ$, with all directions from the back of the mountain (the 
x-axis is pointing to the East). 
The flux of neutrino is assumed to be isotropic and uniform in the field 
of view of the detector. 
Every incident $\nu_\tau$ is tested to see if it 
interacts inside the rock. The energy and momentum of the produced $\tau$ are 
traced until its decay in the case that the neutrino interacted. The energy 
loss and the range of $\tau$ lepton are calculated according to the result of 
Ref.\,~\cite{duttaina}. 
If the $\tau$ decays outside the rock, there is about an 80\% probability that 
an electron or multiple hadrons, mainly pions or kaons, are produced in the decay. 
Those particles will initiate electromagnetic or hadronic showers 
from the decay point. The shower type and energy are determined by a standard 
$\tau$ decay routine and passed to an air shower generator.
However, if the $\tau$ flies too far from the mountain surface before 
its decay, showers could be initiated behind the detector and would not be 
detected. In the simulation, $\tau$'s that decay anywhere further than 
7 km from the escape point on the mountain surface are ignored.

On the other hand, if the $\tau$ decays inside the rock, a regeneration 
procedure will be started in the simulation. This procedure repeats the 
previous process using the decay-product $\nu_\tau$ whose energy is 
determined by the same $\tau$-decay routine mentioned before. Since the 
energy reduction is rather 
large in the $\nu_\tau$ to electron/hadron conversion, there is no need to 
repeat this regeneration procedure in further decays because the shower 
energy is lower than the threshold of the detector. The regenerated 
$\tau$'s which decay inside the rock again are ignored.

\subsection{From tau neutrino to tau lepton}
\begin{figure}
  \epsfig{file=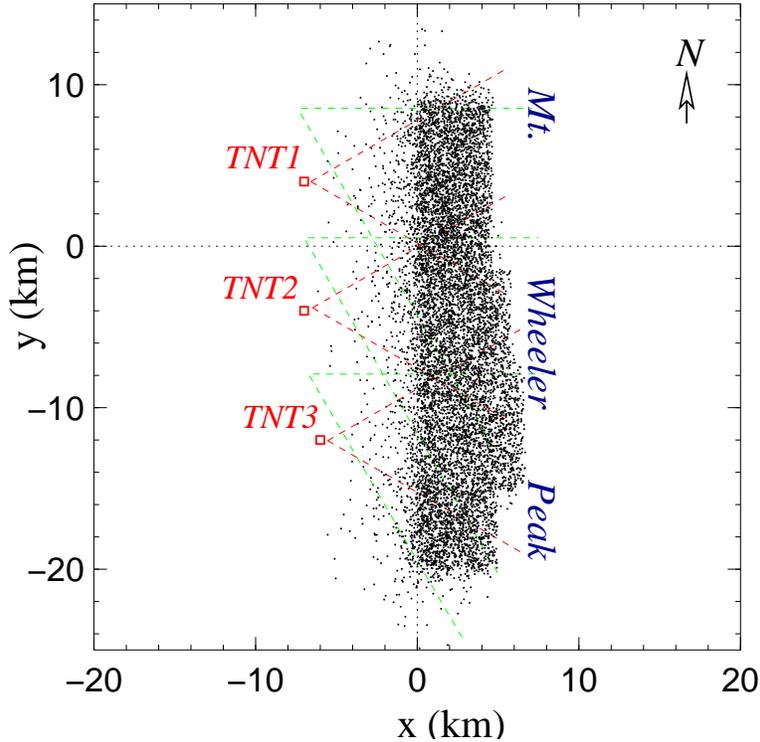,width = 10cm}
  \caption{Locations of initiating points of showers induced by $\tau$ neutrinos 
through MT. Wheeler Peak near Utah/Nevada border. The squares represent CRTNT 
telescopes. The solid lines attached to the detectors indicate the FOV of the 
telescopes. The dashed lines show a further optimization idea for catching 
$\nu_\tau$'s from the center of our galaxy. See text in Conclusion section 
for details.}
  \label{crtnt1}
\end{figure}
The $\nu_\tau$ to shower conversion efficiency is defined as the ratio between 
the total number of successfully converted events and the number of incident 
$\nu_\tau$'s. The simulation yields the conversion efficiency of 
$1.99\times10^{-4}$ and $2.21\times10^{-2}$ for the AGN\cite{AGN} and GZK\cite{GZK} neutrino source 
models, respectively. A distribution of locations where showers start developing
is shown in FIG.\,\ref{crtnt1}. The detectors and their fields of view are shown 
in the same figure. The distributions of trigger efficiency as a function of 
energy are shown in FIG.\,\ref{simu_AGN} and FIG.\,\ref{simu_GZK}. 
The conversion probability is different for different models mainly because 
the GZK neutrino energy spectrum is much harder than the AGN's, therefore 
they have a different pile-up effect associated with $\tau$ energy-loss. 

Because of the rather strong energy loss of the $\tau$ through ionization and radiation, the 
energy of re-converted shower via $\nu$-regeneration is significantly lower. 
The contribution from the $\nu_\tau$-regeneration to the final detectable 
shower event rate is not significant. 
The regenerated $\nu_\tau$-shower conversion efficiency is less than 0.6\% and 
2.0\% of the total conversion efficiency for AGN
and GZK neutrinos, respectively. 
\subsection{ Air Shower Simulation}
\label{corsika}
Corsika 6.0~\cite{CORSIKA} is used to generate air showers in the space 
between shower initiating point outside the mountain and the
CRTNT telescopes. Because the detector is located about 12 
km away from the mountain peak horizontally, the typical distance available for shower 
development is about 10 km in the air. Since all showers are initiated in an 
almost horizontal direction, the air density along the path of the shower 
development is nearly constant. Thus a uniform air density is assumed 
for the air shower development. A simulated shower library is 
established by generating 200 hadronic and 200 electromagnetic showers at 
three values of $1\times$, $2\times$ and $5\times$ of energies in each order 
of magnitude between 10 PeV and 1 EeV. The shower longitudinal profiles, 
i.e. number of charged shower particles along the 
shower axis, are stored in the library at every 5 g/cm$^2$. 

\begin{figure}
  \epsfig{file=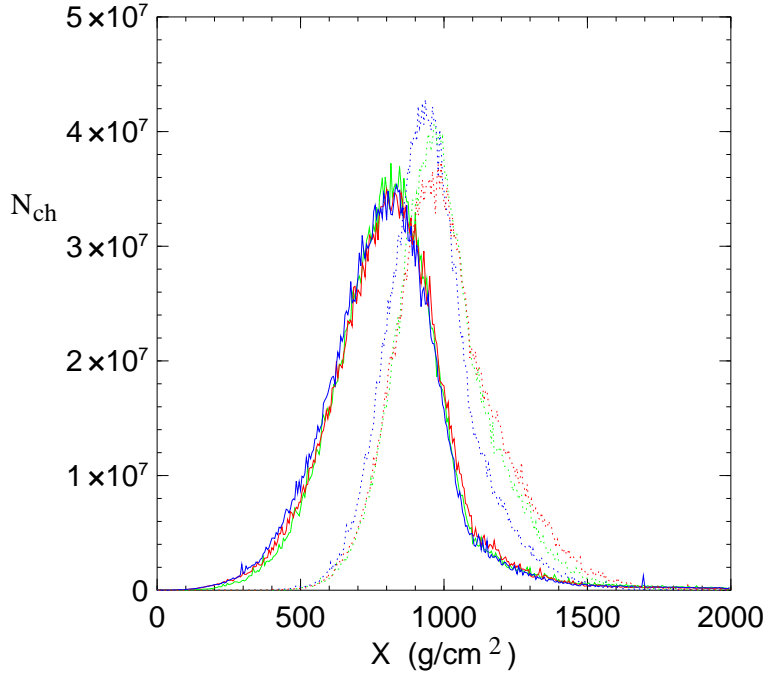,width = 10cm}
  \caption{Showers generated by Corsika at energy 5$\times 10^{16}$ eV. 
Solid lines are hadronic showers. Dotted lines are electron showers.}
  \label{shws}
\end{figure}

Once a $\tau$ decays outside the mountain surface, it starts an air shower  
except if the decay product is a $\mu$. The decay point is 
taken as depth zero g/cm$^2$ for the shower development. A shower profile 
is randomly selected from the library in the group of showers at the 
closest energy. The 
number of charged particles of this picked shower are scaled up or 
down to represent the shower to be simulated. The amount of scaling is 
determined by the difference between the given shower energy and the 
energy of the selected shower. 
Some shower examples are shown in 
FIG.\ref{shws}. 
Note that the longitudinal development behavior is quite different from that of 
showers that develop from the top of the atmosphere to the ground where the 
density of the air increases nearly exponentially. Moreover, the 
hadronic showers develop noticeably faster than electromagnetic showers.

\subsection{Photon production and light propagation}
Charged shower particles excite the nitrogen molecules as they pass through 
the atmosphere. The deexcitation of the molecules generates ultra-violet 
fluorescence light. The number of fluorescence photons is proportional to 
the shower size and these photons are emitted isotropically. The shower 
simulation carried out in this paper assumes a fluorescence light spectrum 
according to a recent summary of world wide measurements\cite{nagano}, 
including the dependence of the yield to 
the atmospheric pressure and temperature. 

Since energies of charged shower particles are higher than the critical energy, 
the shower particles generate Cerenkov photons at every stage of the shower 
development. 
The accumulated Cerenkov light 
concentrates in a small forward cone, therefore the intensity of the light is 
much stronger than the fluorescence light along the shower direction. Even for 
low energy showers at 500 TeV, the strong forward Cerenkov light beam can be 
detected if the shower comes directly toward the detector. For shower coming 
from other directions, a significant part of the Cerenkov light is scattered 
out via Rayleigh and Mie scattering during the whole shower 
development history. The fraction of this light scattered in the direction of the detector can 
also make a noticeable contribution to detector triggering. 

The procedure of Cerenkov light generation and scattering is fully 
simulated. Detailed description 
about the calculation can be found in Ref.~\cite{HRprototype} and 
references therein.

Shower charged particles and therefore fluorescence light photons, spread out 
laterally following the NKG distribution function. The Moliere
radius parameter of the NKG function is about 95 m at about 1500 m a.s.l.
Photons originating from Cerenkov radiation have an exponential lateral 
distribution from the axis of the shower. Therefore, photons coming from a 
shower are spread over a
range of directions around the shower location in the sky due to its 
longitudinal motion and lateral extension. A ray tracing 
procedure is carried out to follow each photon to the 
photo-cathode of PMT's once the photon source is located in the sky. All detector 
responses are considered in the ray tracing procedure, including mirror 
reflectivity, UV filter transmission, quantum efficiency of photo-cathode, 
location-sensitive response function of the photo-cathode and 
optical effects associated with the off-axial and defocus effects. 
The technical parameters of the detector are set to be the  
same as the existing HiRes detector. More details on the detector 
specifications can be found elsewhere\cite{HRprototype}.

Sky noise photons are randomly added in this ray tracing procedure both in 
time and arrival directions. 

The uncertainty associated with the varying weather conditions is negligible 
for the Rayleigh scattering. The Mie scattering due to the aerosols is more 
dependent of the weather condition that drives the aerosol distribution. 
However, for a detector that has an aperture within 6 km, 
the aerosol scattering contribution to the light extinction is close to  
minimum. The uncertainty in the triggering efficiency due to weather 
conditions is thus small. In the simulation, 
an average model~\cite{Lawrence} of aerosol scattering for standard desert 
in west US is employed. 
\section{Predicted event rate}
For the  CRTNT detector, we calculate the 
event rate using the $\nu_\tau$ to air shower conversion algorithm and the 
shower/detector simulation described above. 
Since the primary neutrino energy spectra are different in different production 
models, induced air showers from 
different sources have slightly different detection efficiency. In order to 
illustrate the difference, two extreme cases are simulated in this paper. 
For a model dominated by low energy neutrinos (still above the 
threshold of the CRTNT detector), we use the AGN neutrino source model~\cite{AGN}. 
For a model with many more higher energy neutrinos, we use the GZK neutrino source 
model~\cite{GZK}. We generate 10$^9$ and 10$^7$ trials for AGN and GZK models, 
respectively.
 
\begin{figure}
  \epsfig{file=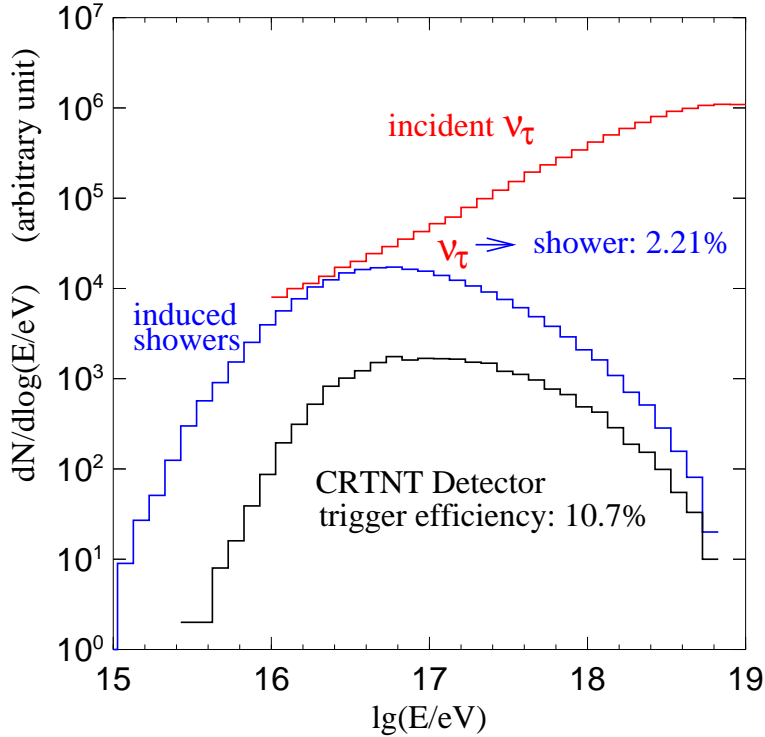,width = 10cm}
  \caption{GZK $\nu_\tau$ to air-shower conversion and triggering rate. The 
incident $\nu_\tau$ energy spectrum, converted shower energy distribution are 
plotted. 
The triggered air-shower event distribution is calculated based on a detector 
described in Sec.\,\ref{detector} and an air-shower development algorithm 
described in Sec.\,\ref{corsika}.}
  \label{simu_GZK}
\end{figure}

Due to the stronger energy loss of the higher energy $\tau$ leptons, 
the observed GZK neutrino 
spectrum is severely distorted. The high energy neutrinos pile up at
 low energies once they are  
converted into lower energy showers. On the other hand, the shower triggering
simulation shows that the trigger efficiency is slightly higher for higher 
energy showers. The competition between those two effects yields a relatively flat 
event rate distribution between 10 PeV and 2 EeV. The distribution is shown 
in FIG.\ref{simu_GZK}. The overall 
detection efficiency is 2.4$\times 10^{-3}$. According to the flux suggested  
by authors of~\cite{GZK} and a typical 10\% duty cycle of the 
fluorescence/Cerenkov light detector, the event rate is about 
$0.23 \pm 0.01$ per year.

The predicted AGN neutrino flux in Ref. \cite{AGN_ss} is ruled out by the 
AMANDA experiment \cite{Amanda}.
In this paper, we use an updated prediction of the AGN neutrino spectrum \cite{AGN}.    
The source spectrum changes from $\sim E^{-1}$ to $\sim E^{-3}$ 
near 10 PeV. The neutrino flux is cut off around 0.6 EeV. I this case, the energy loss of 
the $\tau$ lepton inside the mountain is no longer significant. 
The converted shower 
spectrum has a similar shape to the incident neutrino spectrum at high energies. 
The conversion efficiency drops rapidly with energy in the low energy region. 
This softens the event rate as a function of energy below 10 PeV. 

\begin{figure}  
  \epsfig{file=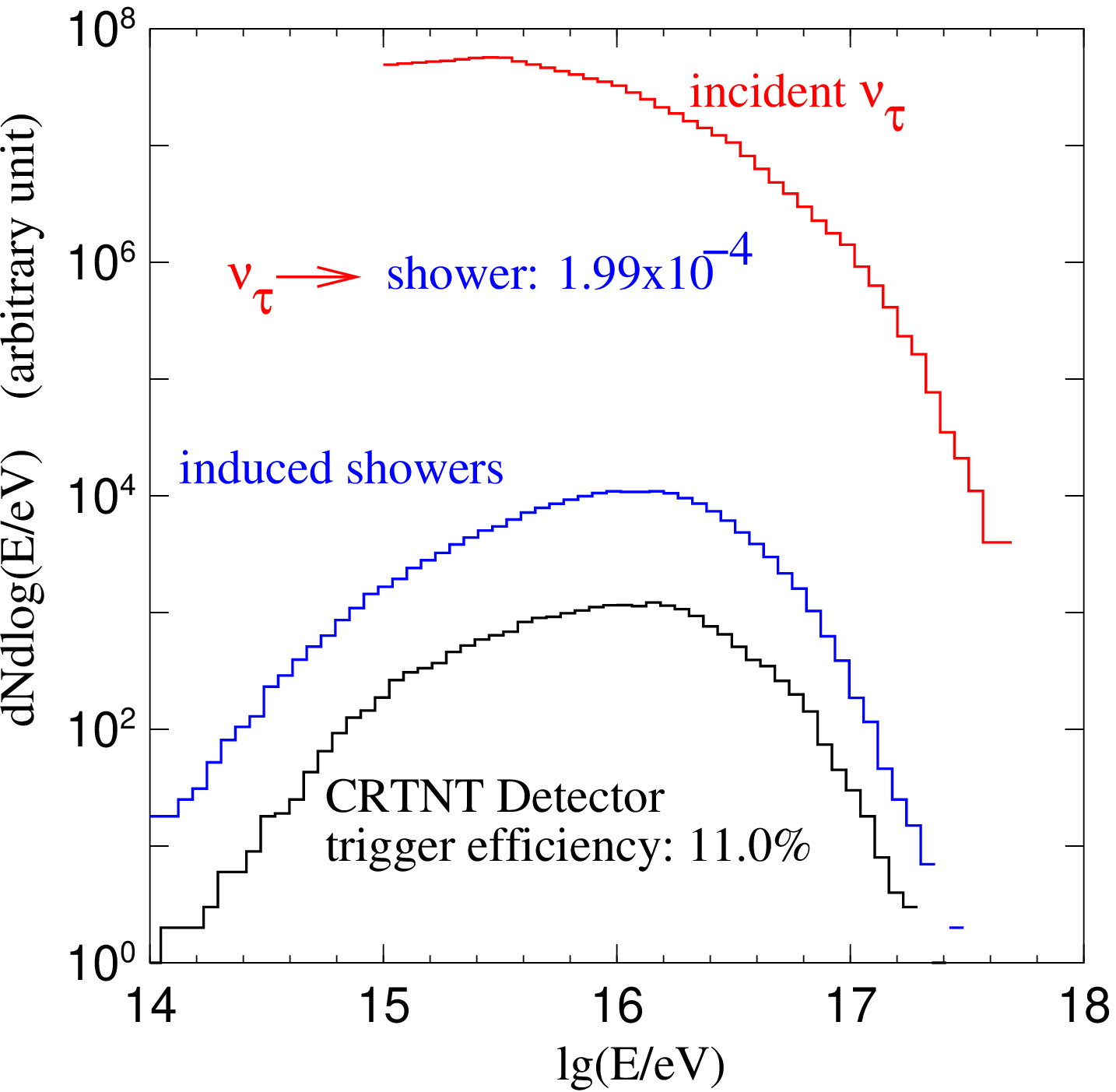,width = 10cm}
  \caption{AGN $\nu_\tau$ to air-shower conversion and triggering rate. The 
incident $\nu_\tau$ energy spectrum and converted shower energy distribution are 
plotted.  
The triggered air-shower event distribution is calculated according to a detector 
described in Sec.\,\ref{detector} and an air-shower development algorithm 
described in Sec.\,\ref{corsika}.}
  \label{simu_AGN}
\end{figure}

The average trigger efficiency of showers induced by the products of 
$\tau$-decays is 11.0\%. The overall detection efficiency of AGN neutrinos 
is $2.19\times 10^{-5}$. The spectra of $\nu_\tau$ are shown in 
FIG.~\ref{simu_AGN}. According to the flux predicted by authors of 
~\cite{AGN}, the event rate is $5.04 \pm 0.05$ per year, 
where a 10\% duty cycle is assumed for the detector.

\subsection{Optimization}
Using the simulation tool, we have tried to optimize the configuration of the 
detector to maximize the event rate. We find that the distances between 
the telescopes and the mountain are important parameters. Without losing 
much of the shadow size, the detection efficiency is 
higher if the telescopes are placed further from the mountain up to the distances
of about 12 km. For the site near Mt. Wheeler Peak, the altitude of the 
detectors are about 1500 m a.s.l. at 12 km from the mountain. This yields
about a 11.7$^\circ$ shadow on the detector. A single layer detector 
configuration shown in FIG.\ref{events} maximizes the trigger efficiency of
the neutrino detection. Using an array configuration of three single layer 
detectors separated by 8 km (see FIG.\ref{crtnt1}) maximizes the usage of 
the mountain body as a $\nu_\tau$ to shower converter.

Since the detector has a limited field of view toward the east, more than 
3/4 of field of view of telescope is covered by the 
mountain shadow. The moon will not be in the field 
of view most of night time. The level of scattered moon 
light toward the detector will be sufficiently low when the moon moves 
behind of the telescopes. The detector should be able to be  
operated under such condition. Conservatively,
the useful observation time can be then increased by 50\%. 
A rate of about 8 events per year could then be expected for the AGN model case.
More accurate estimate requires an on-site measurement for the light background 
with the moon up. 

In this paper, $\nu_e$'s and $\nu_\mu$'s are not taken into account. From the 
point of view of $\tau$ neutrino search, showers induced by these neutrinos 
are background. However, for the purpose of particle astronomy, these 
neutrinos carry the same information about the source as the $\nu_\tau$ does. 
Since the $\mu$ generated in $\nu_\mu + A$ interaction hardly interacts and 
does not initiates a shower in the air, there is nearly no chance for it to trigger the 
detector. However, both $\nu_e$ and $\nu_\mu$ transfer 20\% to 30\% of energy 
to the target in the deep inelastic scattering (DIS) processes. 
The fragments of the target particles can 
induce detectable showers in the air if the neutrinos interact in a very thin 
layer of rock right beneath the surface. The contribution of the DIS processes 
could be enhanced because the neutral current processes also contribute to the 
detectable air shower generation. $\sigma_{CC} + \sigma_{NC}$ is the effective 
DIS cross section, in which the neutral current cross section $\sigma_{NC}$ is 
about 44\% of the charged current cross section $\sigma_{CC}$ at $10^{15}$ eV 
and 31\% at $10^{20}$ eV~\cite{Gandhi}. The interaction length for these 
processes is still very long, e.g. $6.6\times 10^8$ g/cm$^2$ for $10^{16}$ eV 
and $2.2\times 10^7$ g/cm$^2$ for $10^{20}$ eV. Since the total thickness of 
the mountain is about 20 km, i.e. $5.2\times 10^6$ g/cm$^2$, we can assume the 
free paths of neutrinos to be a uniform distribution. Assuming that only DIS's 
which happened within a layer of one to two radiation length 
behind the mountain surface are able to generate detectable showers, the 
DIS event conversion efficiency is approximately $5\times 10^{-6}$. Comparing the 
$\nu_\tau$-shower conversion efficiency approximately $3\times 10^{-4}$ at 
$10^{16}$ eV, the ratio of DIS events versus $\nu_\tau$ events is only about 1.7\%.

       For higher energies, the mean free path of the neutrino is shorter but only 
by one order of magnitude over three decades of energies. Thus the flat 
free path distribution is still a good approximation. The $\nu_\tau$-shower conversion
efficiency increases with energy rather rapidly. The ratio between number of DIS events 
and number of $\nu_\tau$ events reduces to 0.04\% at 10$^{19}$ eV. 
\section{ Conclusion}
The technique of using a mountain to convert $10^{16}\sim 10^{19}$ eV neutrinos
to air showers and using fluorescence/Cerenkov light to detect the showers can be 
optimized by setting the detector near a $\sim$20 km thick mountain body. 
Mt. Wheeler near the Nevada/Utah border is a good example of maximizing the conversion in 
this energy range. The energy reduction in the conversion process is so severe 
for high energy $\tau$'s that  showers pile up below $10^{17}$ eV in the 
shower energy spectrum.  The regeneration of $\nu_\tau$ is found to have an
insignificant effect on the conversion of $\nu_\tau$ to air shower, a
contribution of less than two percent. Cerenkov light may dominate 
over fluorescence light and become the main light source to trigger the CRTNT
 detectors. An air shower development algorithm has 
been developed using Corsika in a uniform atmosphere for estimating trigger 
efficiency. 
Fluorescence/Cerenkov light production and propagation are fully simulated. Detector
response is also fully simulated. The detector configuration is optimized using
this simulation tool.
The CRTNT detector simulation indicates that a rate of about 8 events 
per year for an optimized detector with 15\% duty cycle is expected for the AGN 
source model\cite{AGN}. 

With such a sensitivity, the CRTNT detector can be used for searching for
nearby cosmic ray sources, e.g. our galactic center (GC). Since gamma rays are
blocked by dust near the GC and cosmic rays are bent away from the
GC due to the strong magnetic field around the GC, the neutrino is a unique
particle can be used for exploring the GC region. At the site near Mt.
Wheeler Peak, the GC rises above the horizon in summer night and reaches
its highest elevation about 20 $^\circ$. The GC would be well covered by
the CRTNT detector if the detectors face south. Therefore, to make a
balance between the use of the mountain as a converter and the duration
in which the GC being seen, we turn all telescopes about 30 $^\circ$ to
the south. The field of view of the reconfigured CRTNT detector is
shown in FIG.\ref{crtnt1} by the dashed lines. The neutrino event trigger
efficiency  would not be changed much, however, about 1000 hour exposure
could be expected during a run of three years.

\section{Acknowledgments}
MAH and ZC are grateful to G.L. Lin and J.\,J.\,Tseng for theoretical support. 
Many thanks from ZC to R.W. Springer for very fruitful discussions. 
This work (ZC) is partially supported by Innovation fund (U-526) of IHEP, 
China and Hundred Talents \& Outstanding Young Scientists Abroad Program (U-610) 
of IHEP, China. ZC and PS are partially supported by US NSF grants PHY 9974537 and 
PHY 9904048, while MAH is supported by Taiwan NSC grant 92-2119-M-239-001.

%\section{References} 

%\vspace{\baselineskip}

\end{document}